RESEARCH ARTICLE

# Synaptic polarity and sign-balance prediction using gene expression data in the *Caenorhabditis elegans* chemical synapse neuronal connectome network

Bánk G. Fenyves[1,2], Gábor S. Szilágyi[1], Zsolt Vassy[1], Csaba Sőti[1], Peter Csermely[1]*

1 Department of Molecular Biology, Semmelweis University, Budapest, Hungary, 2 Department of Emergency Medicine, Semmelweis University, Budapest, Hungary

* csermelynet@gmail.com

## Abstract

Graph theoretical analyses of nervous systems usually omit the aspect of connection polarity, due to data insufficiency. The chemical synapse network of *Caenorhabditis elegans* is a well-reconstructed directed network, but the signs of its connections are yet to be elucidated. Here, we present the gene expression-based sign prediction of the ionotropic chemical synapse connectome of *C. elegans* (3,638 connections and 20,589 synapses total), incorporating available presynaptic neurotransmitter and postsynaptic receptor gene expression data for three major neurotransmitter systems. We made predictions for more than two-thirds of these chemical synapses and observed an excitatory-inhibitory (E:I) ratio close to 4:1 which was found similar to that observed in many real-world networks. Our open source tool (http://EleganSign.linkgroup.hu) is simple but efficient in predicting polarities by integrating neuronal connectome and gene expression data.

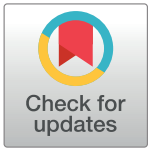

**Data Availability Statement:** All relevant data are within the manuscript and its Supporting information files.

**Funding:** This work was supported by the Hungarian National Research, Development and Innovation Office [Hungarian Scientific Research Fund, K131458 to P.C. and K116525 to C.S.] (https://kormany.hu/emberi-eroforrasok-miniszteriuma), by the Higher Education Institutional Excellence Programme of the Ministry

## Author summary

The fundamental way neurons communicate is by activating or inhibiting each other via synapses. The balance between the two is crucial for the optimal functioning of a nervous system. However, whole-brain synaptic polarity information is unavailable for any species and experimental validation is challenging. The roundworm *Caenorhabditis elegans* possesses a fully mapped connectome with an emerging gene expression profile of its 302 neurons. Based on the consideration that the polarity of a synapse can be determined by the neurotransmitter(s) expressed in the presynaptic neuron and the receptors expressed in the postsynaptic neuron, we conceptualized and created a tool that predicts synaptic polarities based on connectivity and gene expression information. Using currently available datasets we propose for the first time that the ratio of excitatory and inhibitory synapses in a partial connectome of *C. elegans* is around 4 to 1 which is in line with the balance observed in many natural systems. Our method opens a way to include spatial and temporal dynamics of synaptic polarity that would add a new dimension of plasticity in the

excitatory:inhibitory balance. Our tool is freely available to be used on any network accompanied by any expression atlas.

## Introduction

Chemical synapses of a neuronal network are both directed and signed, since a neuron is able to excite or inhibit another neuron. The nervous system of the nematode *Caenorhabditis elegans* has been fully mapped and reconstructed [1–3]. However, except for a few connections there is no comprehensive chemical synapse polarity data available [1]. While the direction of a synaptic connection can be inferred from its structure, the experimental determination of its polarity requires delicate electrophysiological methods with limited system-level use (e.g. patch-clamping) or more recent calcium-imaging or optogenetic techniques. Instead, *in silico* approaches using reverse engineering and genetic algorithms have efficiently predicted synaptic sign for different subnetworks of the *C. elegans* connectome [4–8].

Many synaptic sign prediction models have relied on the widely accepted assumption that the polarity of a chemical synapse is solely determined by the type of neurotransmitter released by the presynaptic neuron [4,5]. Therefore, in *C. elegans* excitatory glutamatergic and cholinergic, as well as inhibitory γ-aminobutyric acid (GABA)-ergic ionotropic chemical connections have been modeled. However, with this approach, approximately 6% of the connections turned to be inhibitory [9,10]. A low proportion of inhibitory connections can result in an unbalanced, over-excited network, as has been shown by previous publications [11–13]. Moreover, there is evidence of unconventional postsynaptic effects of neurotransmitters, such as cholinergic inhibition [14,15] or glutamatergic inhibition [16–18], meaning that a neuron can simultaneously excite and inhibit its postsynaptic partners with the same neurotransmitter due to variable neurotransmitter receptor expression on the postsynaptic neuron membrane. For example, the cholinergic AIY interneuron can activate RIB neurons and inhibit AIZ neurons in an acetylcholine-mediated fashion [15].

We aimed to predict synaptic polarities in the *C. elegans* ionotropic chemical synapse connectome (297 neurons and 20,589 synapses) relying on presynaptic neurotransmitter and postsynaptic receptor gene expression data for the three main neurotransmitters glutamate, acetylcholine and GABA. In this study of the *C. elegans* nervous system, we predicted the polarity of more than 70% of ionotropic chemical synapses and predicted a sign-balance of excitatory:inhibitory connections close to 4:1 that has been observed as functionally stable in many real-world circumstances. Presenting a new dataset, we show that the concept of gene expression-based polarity prediction can efficiently be applied to demonstrate a balanced E:I ratio in a nervous system.

## Results

### Creating a prediction tool of the synaptic polarities of the *C. elegans* connectome

Our primary goal was to infer synaptic polarity from combining connectivity and gene expression data. We created a simple, yet powerful algorithmic database (S1 Data) that takes as input connectome and gene expression data to predict synaptic polarity of ionotropic glutamatergic, cholinergic and GABA-ergic connections. We used the *C. elegans* WormWiring connectome data primarily in the form of a weighted edge list representing 20,589 chemical synapses in 3,638 connections. Our prediction tool is available here: http://EleganSign.linkgroup.hu.

### Update of the previous neurotransmitter expression tables

We updated the *C. elegans* neuronal neurotransmitter tables previously published [19–22] with recent evidence [10,23] (Methods). After this update, 256 neurons had a single neurotransmitter expressed, while 12 neurons had double neurotransmitter expression (Fig 1A). There were 34 neurons which did not express any of the three neurotransmitters investigated.

### Extraction of gene expression data

In parallel, we extracted gene expression data from Wormatlas [24], Wormbase (www.wormbase.org), and a recent RNA-sequencing dataset [23] which we curated manually to assign ionotropic

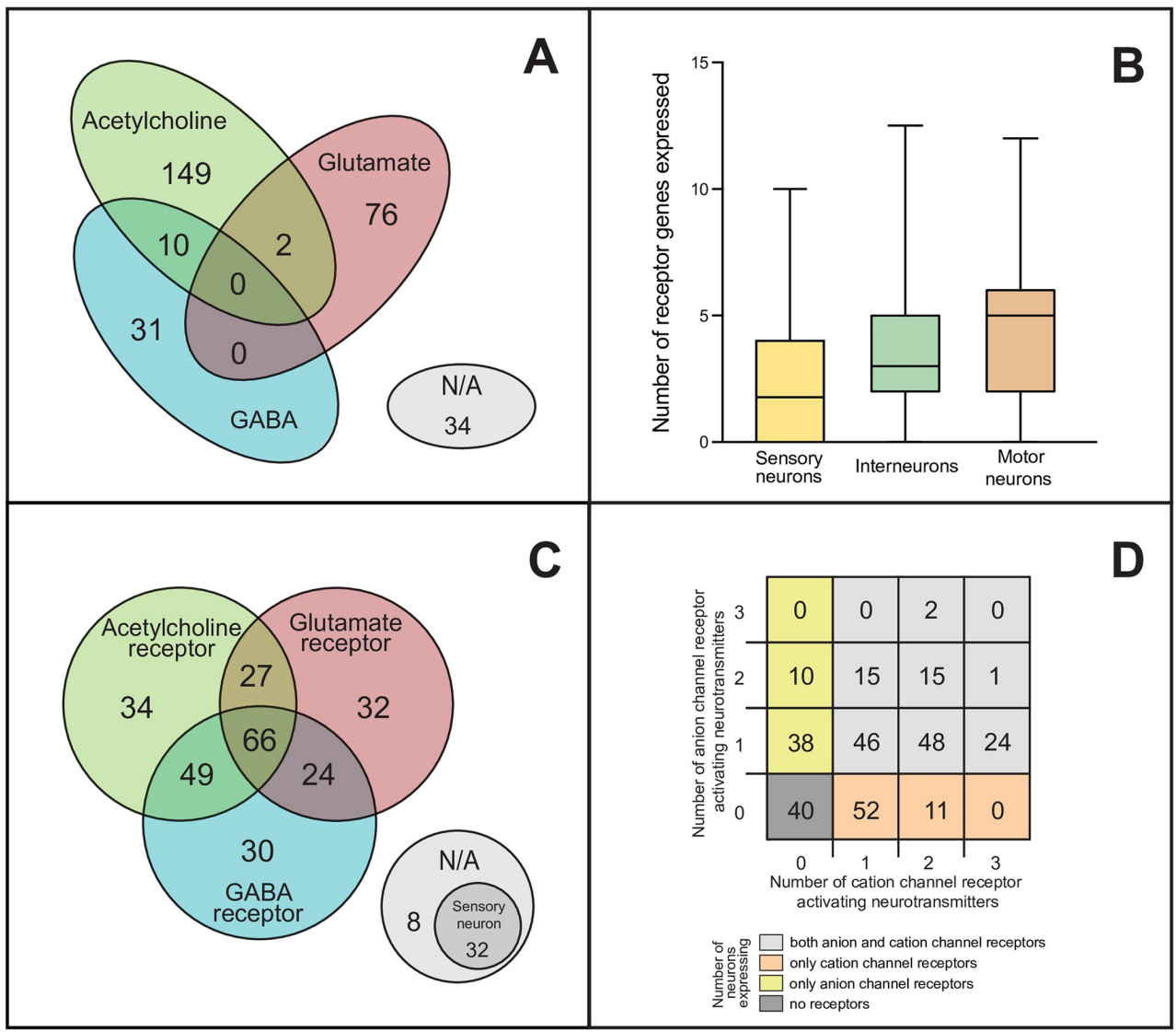

**Fig 1. Neurotransmitter and receptor expression patterns of *C. elegans* neurons.** Expression data of the three major synaptic neurotransmitters and their receptors of *C. elegans* were collected from multiple datasets and were manually curated (see Methods). **(A)** Distribution of neurons according to their neurotransmitter expression: glutamate (red), acetylcholine (green), GABA (blue) or none (grey). **(B)** Number of receptor genes expressed by neurons, grouped by neuron modality. **(C)** Distribution of neurons based on their neurotransmitter receptor gene expression (colors are the same as in panel A). **(D)** Distribution of neurons according to the number of neurotransmitters for which anion and/or cation channel receptor genes are expressed.

https://doi.org/10.1371/journal.pcbi.1007974.g001

receptor expression pattern to each neuron (see Methods). To do this we first sorted the previously identified 62 ionotropic receptor genes into six functional classes based on their suggested neurotransmitter ligand (glutamate, acetylcholine or GABA) and putative ion channel type (cationic or anionic, i.e. excitatory or inhibitory), as shown in Table 1. We found evidence for postsynaptic neuronal expression of 42 out of the 62 receptor genes in the *C. elegans* nervous system (genes marked bold in Table 1; also S1 Data). We also found an increasing average number of receptor genes in sensory, inter- and motor neurons, respectively (Fig 1B).

Next, for all the 302 neurons of the *C. elegans* connectome we determined which receptor classes were expressed. 166 neurons had an overlapping expression of receptors for two or three different neurotransmitters (Fig 1C). The distribution of neurons according to their expression of cationic and/or anionic glutamate, acetylcholine and/or GABA receptors suggested functional diversity due to the high number of neurons expressing both excitatory and inhibitory receptors (Fig 1D). Surprisingly, 85 neurons expressed both excitatory and inhibitory receptor genes for the same neurotransmitter (S1 Data). Forty out of 302 neurons showed no receptor expression, out of which 32 neurons were primarily sensory neurons (S1 Data). The average number of receptor genes expressed was 3.7 per neuron (S2 Table).

## Neurotransmitter and receptor gene expression-based polarity prediction

After assigning neurotransmitter and receptor expression patterns to each neuron, we predicted synaptic polarities by looking for matches between the neurotransmitter expression of the presynaptic neuron and the receptor gene expression of the postsynaptic neuron (Fig 2A). This way, we labeled synapses as one of the following: excitatory, inhibitory, complex, or

**Table 1. Neurotransmitter receptor genes.**

| | Glutamate | Acetylcholine | | GABA |
|---|---|---|---|---|
| **Cation channel receptor gene** | *glr-1* | *acr-1* | *acr-16* | *exp-1* |
| | *glr-2* | *acr-2* | *acr-17* | *lgc-35* |
| | *glr-3* | *acr-3* | *acr-18* | |
| | *glr-4* | *acr-4* | *acr-19* | |
| | *glr-5* | *acr-5* | *acr-20* | |
| | *glr-6* | *acr-6* | *acr-21* | |
| | *glr-7* | *acr-7* | *acr-23* | |
| | *glr-8* | *acr-8* | *acr-25* | |
| | *nmr-1* | *acr-9* | *deg-3* | |
| | *nmr-2* | *acr-10* | *des-2* | |
| | | *acr-11* | *eat-2* | |
| | | *acr-12* | *lev-8* | |
| | | *acr-13* | *unc-29* | |
| | | *acr-14* | *unc-38* | |
| | | *acr-15* | *unc-63* | |
| **Anion channel receptor gene** | *glc-1* | *acc-1* | | *gab-1* |
| | *glc-2* | *acc-2* | | *ggr-1* |
| | *glc-3* | *acc-3* | | *ggr-2* |
| | *glc-4* | *lgc-47* | | *ggr-3* |
| | *avr-14* | *lgc-48* | | *lgc-36* |
| | *avr-15* | *lgc-49* | | *lgc-37* |
| | | | | *lgc-38* |
| | | | | *unc-49* |

The *Caenorhabditis elegans* genome contains 62 ionotropic postsynaptic receptor genes for glutamate, acetylcholine, and GABA. *acc-4* and *lgc-46* genes were excluded from our database due to suggested presynaptic expression (S1 Table). In this table genes are grouped according to their neurotransmitter ligand and whether forming cationic (+) or anionic (−) ion channels (based on [24] and other references listed in S1 Table). In *C. elegans* "unconventional signaling", namely, glutamate-mediated inhibition, cholinergic inhibition and GABA-ergic excitation, is facilitated by 6, 6, and 2 receptor genes, respectively. In the gene expression database used in this work, expression in at least one neuron was found in the case of 42 genes (marked **bold**), while for 20 genes no neuronal expression was found.

https://doi.org/10.1371/journal.pcbi.1007974.t001

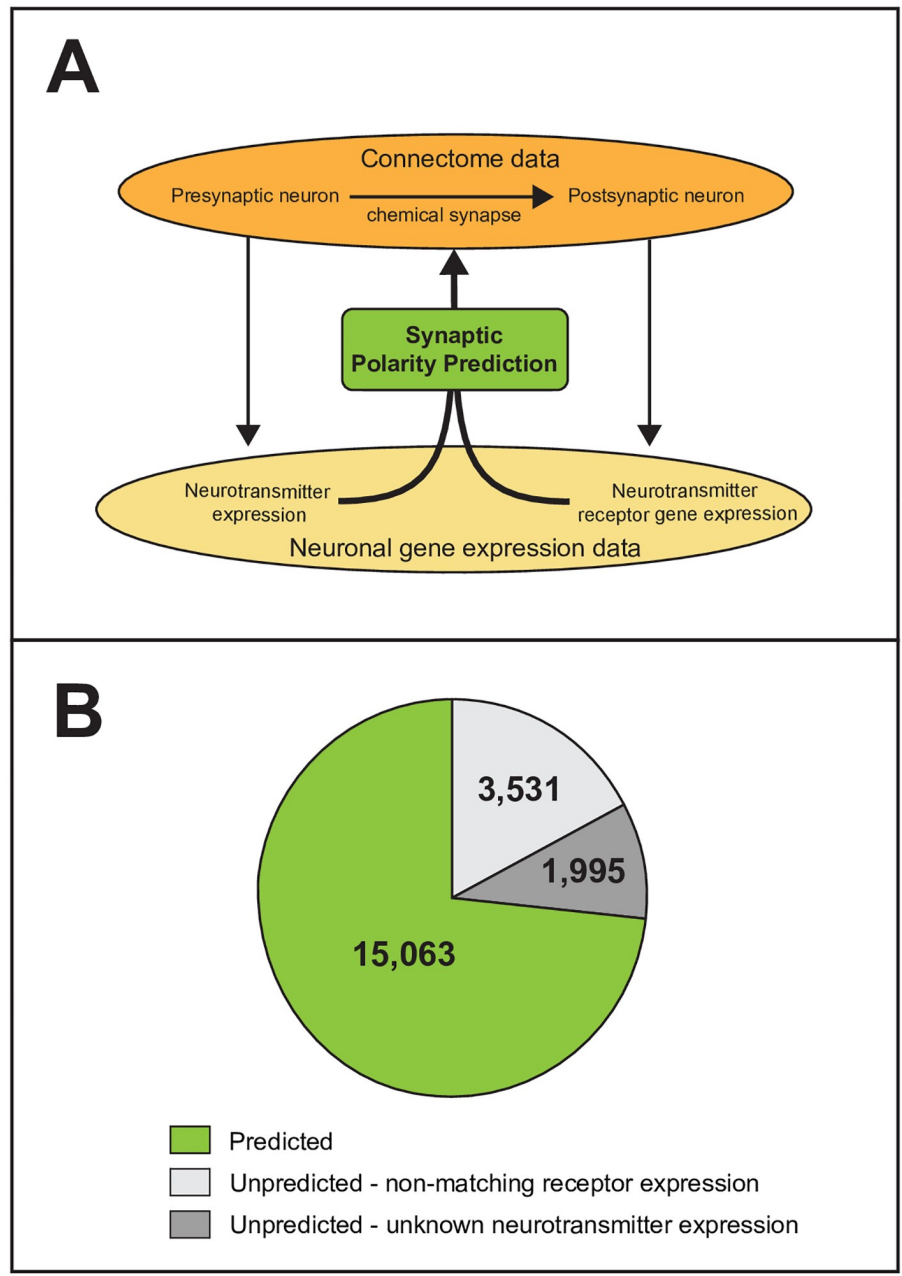

**Fig 2. Prediction of synaptic polarities of the *C. elegans* ionotropic chemical synapse connectome. (A)** Prediction method. Connectome and gene expression data were manually curated (see Methods). Polarities of chemical synapses were predicted based on the neurotransmitter expression of presynaptic neurons and the matching receptor gene expression of the postsynaptic neurons. **(B)** Distribution of predicted and unpredicted synapses. We were able to predict polarity for 73% of chemical synapses (green). The polarities of the rest of synapses were unpredicted due to unknown neurotransmitter expression of the presynaptic neurons (dark grey) or non-matching receptor gene expression of the postsynaptic neurons (light grey).

unpredicted (see Methods). "Excitatory" or "inhibitory" label were given when the neurotransmitter-matched postsynaptic receptor genes were only cation or anion channel related, respectively. A synapse was labeled as "complex" if data suggested both excitatory and inhibitory function. With this approach, we predicted synaptic polarity for 73% of chemical synapses of the *C. elegans* connectome (Fig 2B). We could not predict polarity for the remaining synapses due to missing neurotransmitter data or mismatch in neurotransmitter/receptor expression (Fig 2B). We predicted that 9,070 of the synapses are excitatory and 2,580 are inhibitory, while 3,413 synapses have complex function (Fig 3A and S1 Data). These findings suggest that the overall ratio of excitatory and inhibitory synapses (E:I ratio) in the *C. elegans* ionotropic chemical synapse network is close to 4:1 (Fig 3A, *NT+R method*).

## Alternative polarity prediction methods

To put our results in context, we applied two alternative prediction methods for comparison (*NT-only* and *R-only*; see Methods). The *NT-only* method yielded a much higher E:I ratio

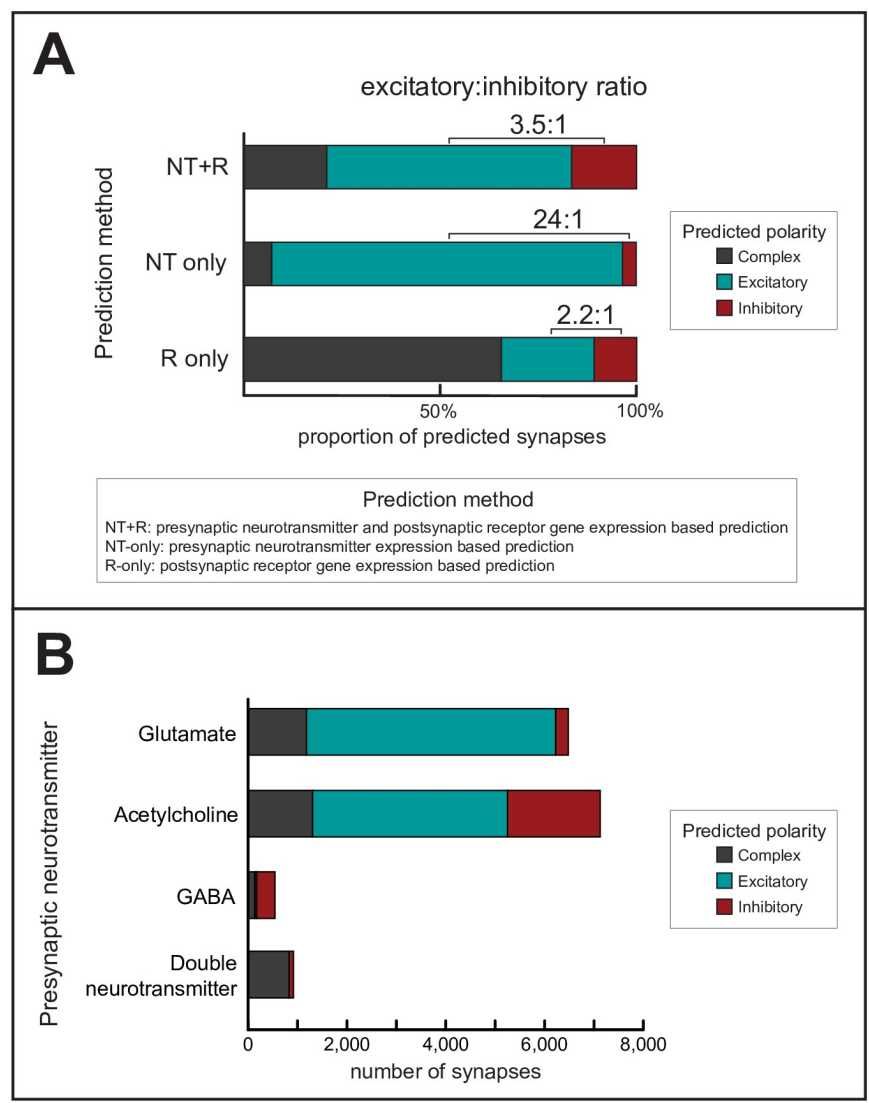

**Fig 3. Predicted synaptic polarities. (A)** Distributions of predicted polarities, using the method developed in this paper (*NT+R*) and two alternative methods as comparison (*NT-only* and *R-only*). Polarities were predicted by considering the neurotransmitter expression of the presynaptic neuron and/or the receptor gene expression of the postsynaptic neuron (see Methods). **(B)** Distributions of predicted synaptic polarities (using the *NT+R method*) according to the presynaptic neurotransmitter. Colors are the same as in panel **A**. Unpredicted synapses are not shown.

https://doi.org/10.1371/journal.pcbi.1007974.g003

(Fig 3A, *NT-only method*; S2 Data), which is in line with the dominance of purely glutamatergic or cholinergic (traditionally excitatory) neurons over GABA-ergic (traditionally inhibitory) neurons (Fig 1A). To explain the difference further, the *NT+R* method predicted that 30% of cholinergic and 5% of glutamatergic synapses were *inhibitory* (Fig 3B) which is a significant fraction of synapses otherwise predicted excitatory with the *NT-only* method. A pairwise comparison of polarities predicted with the *NT+R* and *NT-only* methods is presented in S2 Data. The other, *R-only* method yielded a markedly smaller E:I ratio, however predicted an excessive number of complex synapses (Fig 3A, *R-only method*, S3 Data). This is due to the fact that many neurons express both cation and anion channel receptor genes (Fig 1D).

### Feedback inhibition between neuron groups

Notably, in subsets of connections which connect neurons of different modalities of sensory neurons, motor neurons, interneurons and polymodal neurons, the E:I ratios varied between 1:10 (motor-> sensory) and 14:1 (inter-> motor). Importantly, we observed dominant inhibition in the motor-> sensory, motor-> inter, and inter-> sensory directions (Fig 4A), exhibiting inhibitory backward signaling as discussed in the literature previously [25,26], as opposed to a forward (sensory -> motor) excitatory excess. Besides, a significant presence of inhibitory and complex connections was found in the locomotion circuit as well (Fig 4B and S2 Fig).

### Network representations of the signed *C. elegans* connectome

Network representations of synaptic polarities in the *C. elegans* ionotropic chemical synapse connectome using the EntOptLayout plugin of Cytoscape [27] are in Fig 5. Fig 5A shows that the modular structure of the *C. elegans* connectome visualized by the EntOptLayout method nicely captures the anatomical locations of the anterior, ventral and lateral ganglions, as well as the premotor interneurons of the worm. While the anterior and lateral ganglions show a large glutamate expression, this is much less characteristic to the ventral ganglion (Fig 5A). Fig 5B shows that the ventral ganglion has predominantly inhibitory connections, while connections in the other locations are predominantly excitatory if predicted by our *NT + R method*. Fig 5C demonstrates that the prediction of polarities by neurotransmitters only (*NT-only method*) results in a large excitatory excess, mainly because the connections predicted as inhibitory or complex with the *NT+R method* turn into excitatory. This difference can be observed amongst head neurons and premotor interneurons, but less amongst motor neurons. Many connections of the polymodal neurons are predicted as complex only with the *NT+R method*. The anatomical locations of neurons expressing various neurotransmitters (Fig 5D) correspond well to the network representation shown in Fig 5A. Links between inhibitory function and anatomical structures have been shown in the human brain [28,29], but have not been demonstrated previously in the nematode.

### Validation of our predictions

To validate our results, we contrasted our predictions to previously published *in silico* and experimental work as well. Comparison to the computational findings of Rakowski & Karbowski [30] in the *C. elegans* locomotion circuit of 7 neuron groups and 652 synapses showed a 70% consistency in predicted synaptic polarities (53% on the level of connections) (S3 Table), albeit using a completely different concept. When testing our predictions against experimental evidence based on the literature we found that the majority of predicted polarities using our method were consistent with earlier findings in *C. elegans*: only 1 out of 12 interneuronal connections (29 / 501 synapses) reviewed was predicted an opposing polarity to what has

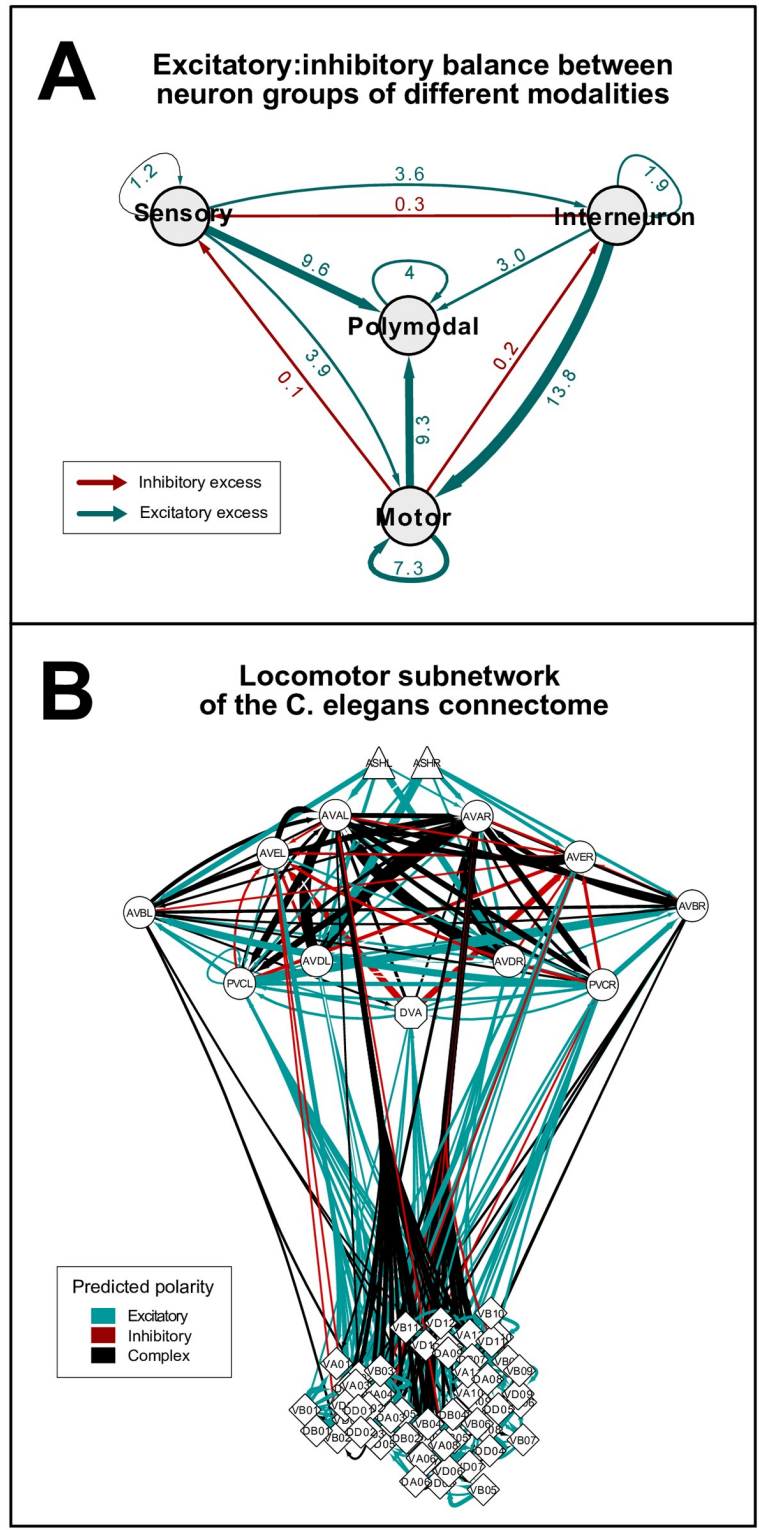

**Fig 4. Excitatory:inhibitory balance of different neuron groups. (A)** Excitatory:inhibitory balance between neuron groups of different modalities. Nodes represent groups of neurons by modality. Edges are weighted according to the excitatory:inhibitory (E:I) ratios (see numbers). Green and red colors represent excitatory (E:I>1) and inhibitory (E: I<1) excess in sign-balances, respectively. **(B)** Network representation of the locomotion subnetwork. Edges represent excitatory (blue), inhibitory (red), or complex (black) chemical connections. Edges are weighted according to synapse number. The shape of vertices (Δ,◦,⋄) represent the modality (sensory, inter, motor, respectively) of neurons. Separate representations of the head circuit and the ventral nerve cord motor neurons are in S2 Fig.

https://doi.org/10.1371/journal.pcbi.1007974.g004

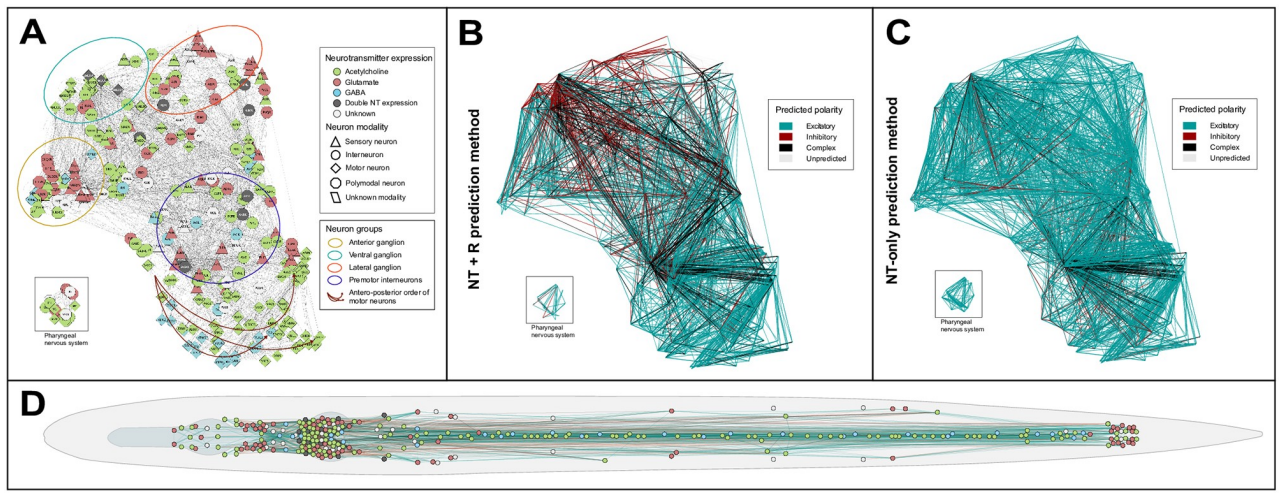

**Fig 5. Network representations of the *C. elegans* chemical synapse network. (A-C)** Network representations using the EntOpt layout plugin in Cytoscape [27]. **(A)** Color and shape of vertices represent neurotransmitter expression and modality of neurons, respectively (see inset for definitions). **(B)** Edges represent excitatory (blue), inhibitory (red), or complex (black) chemical connections predicted by the NT+R method (see Methods), weighted according to synapse numbers. **(C)** Colors of edges (see panel **B**) represent the polarities of chemical synapses predicted by the NT-only method. **(D)** Layout of vertices is representing the anatomic position of neurons. Node and edge colors are as in panels **A** and **B**, respectively. High-resolution representation is available in S1 File.

https://doi.org/10.1371/journal.pcbi.1007974.g005

experimentally been confirmed (S4 Table). This ratio is 6 /12 when validating the *NT-only* prediction method, supporting the importance of receptor expression.

To test the robustness of our *NT+R* prediction method to predict E:I balance, we applied the same rules to predict polarities of connections regardless of synapse number data, and after perturbations in the network like deletion of the 20 pharyngeal nervous system neurons or deletion of potentially variable (i.e. single-synapse) connections (S4 Data). Furthermore, we repeated our analyses in other published connectomes [2,3] of different sizes as well (S5 Table and S5 Data). In all five cases, the excitatory:inhibitory ratios were in the range of 3.1 to 4.1 (S1 Fig). When predicting not using the yet preprint-published RNA-seq expression dataset [23], this range was 3.7–4.1 (S6 Data). All together, these findings suggest that the observed sign-balance is a remarkably robust property of the *C. elegans* ionotropic chemical synapse network.

## Discussion

Nervous systems are not only directed but signed networks as well since neurons either activate or inhibit other neurons [12]. The balance of excitatory and inhibitory connections (i.e. the sign-balance) is a fundamental feature of brain networks, clearly marked by the variety of disorders associated with its impairment [12,31,32]. However, direct evidence of single synapse polarity is rather sporadic even in simple species. In this work we predicted that the sign-balance in the *C. elegans* ligand-gated ionotropic chemical synapse network is approximately 4:1 (excitatory-inhibitory, E:I). This is consistent with previous *in vitro* and *in vivo* studies of nervous systems [33–36], and also with observations of different social networks [37,38], as shown in Table 2. However, this ratio can only be predicted if not only the neurotransmitter expression of the presynaptic neuron but also the receptor gene expression of the postsynaptic neuron is taken into consideration. Its significance is due to fourteen receptor genes that are presumed to encode inhibitory glutamatergic/cholinergic or excitatory GABA-ergic postsynaptic ion channel receptors. This concept of unconventional signaling is not new, but has already been described in *C. elegans* [14,39,40] and other primitive species [41–43], and also in

**Table 2. Proportions of negative edges in signed networks.**

| Network | Proportion of negative edges | Reference |
|---|---|---|
| *C. elegans* chemical synapse network | 20% | this work |
| Rat hippocampus (*in vivo*) | 5–30% | [35] |
| Rat excitatory neocortical neurons | 20% | [36] |
| Cerebral cortex (*in vivo*; GAD expression) | 10–20% | [17] |
| Cerebral cortex (*in vivo*; GABA neurons) | 20–25% | [17] |
| Optimal network for synchronized bursting activity (*in silico*) | 10–20% | [6] |
| Primary visual cortex (V1; *in silico*) | 25% | [49] |
| Neuronal network (*ex vivo*) and neuronal network model (*in silico*) | 20% | [18] |
| Wikipedia (social network) | 21% | [37] |
| Epinions (social network) | 15% | [37] |
| Slashdot (social network) | 23% | [37] |
| University freshman network (social) | 12–14% | [38] |

https://doi.org/10.1371/journal.pcbi.1007974.t002

mammals in the postnatal period [44,45]. This concept has already motivated the prediction of connection polarity instead of neuron polarity, yet on a subcircuit level [30]. Complementing recent work that used gene expression data for structural and functional modelling of the nervous system [46–48], our prediction model is a novel attempt to predict polarities of all ligand-gated ionotropic chemical synapses of the *C. elegans* connectome.

A surprisingly high proportion of synapses were predicted to have a complex, i.e. both excitatory and inhibitory polarity. This is due to parallel expression of cationic and anionic receptor genes–often for the same neurotransmitter–in half of the neurons. This suggests a highly complex functioning of neuronal connections that extends beyond the permanently exclusive concept of excitation-inhibition dichotomy. Since our work is mostly based on expression data of subunits instead of functional receptors, predictions made are from genetic permissibility rather than direct receptor complex presence. While ionotropic transmission in a single synapse is typically either excitatory or inhibitory, the predicted "complexity" can be resolved mainly in two physiologically well-established ways. One is that postsynaptic receptors are not homogenously distributed across the plasma membrane but their subcellular localization is regulated. This allows the receptors to act independently [50–55] and allows the same neurotransmitter to excite and inhibit at distinct postsynaptic sites. For example, such mechanisms have been identified in the AIA [56–58] and AIB neuron groups [59].

Another explanation of "complexity" is the dynamic change of gene expression in time which is observed all through the life-span of a worm e.g. during development, learning (synaptic plasticity), and aging [60–68]. Ultimately, changes in gene expression can lead to neurotransmitter-switching and consequential up- and downregulation of receptors of opposing polarity [69,70]. In its complexity, co-transmission by parallel expression of different neurotransmitters and receptors is one of the mechanisms of plasticity [71].

Currently, there is not enough data to address either the spatial or the temporal aspects of receptor expression regulation on the worm-scale. As expression-profiling methods will provide whole-brain and dynamic proteomics data of subcellular resolution, complex synapses might be further resolved.

There are several mechanisms of interneuronal communication acting in concert to transmit signals while maintaining a responsive but balanced system. The balance of excitation and inhibition—crucial for network stability—is reached via a number of mechanisms. Both synaptic and extrasynaptic, electrical and chemical, voltage-gated and ligand-gated, ion channel-mediated and G-protein coupled neurotransmission have diverse but intertwining roles in promoting and modulating excitation and inhibition. As there are significant mechanistic and functional differences (e.g. speed, modulatory role) between transmission modes, the neuronal connectome can be comprehended as a multiplex network of partially independent layers with each layer representing a distinct type of signaling [47,72,73]. Two dedicated cases were monoamine and metabotropic transmission which were excluded from our workflow. Since monoamine transmission via serotonin, dopamine and tyramine typically occurs extrasynaptically (i.e. 94% in case of tyramine-responsive neurons), expression data of even ionotropic channels (e.g. *mod-1*, *lgc-55*) would have been difficult to apply on the hard-wired connectome used in our study [47,74,75]. Additionally, in [47] the authors created a wireless (extrasynaptic) connectome of *C. elegans* based on matching monoamine/neuropeptide expression with receptor gene expression showing a network structurally different from the hard-wired connectome. Wireless networks possibly exist for ionotropic receptors as well via mechanism of spillover transmission [10,76]. Likewise, metabotropic neurotransmission typically acting via G-protein coupled receptors plays a rather modulatory than direct excitatory/inhibitory role in the nervous system by inducing broad, long-lasting, slow time-scale changes which is distinctive [71,73,77–80]. Our paper covers ligand-gated ionotropic synaptic connections, which account for the fast-acting system of neurotransmitter-mediated synaptic signaling. Additional layers of neural signaling can be targets of polarity prediction in subsequent studies and potentially overlayed on this signed network.

There are a number of limitations of our study which limit its generalizability at the current state: 1) although the most complete of any species, new connectivity data of the worm is still emerging [3,60] as well as 2) gene expression data [23,81]; 3) although our assumption that cation and anion channels are consistently excitatory and inhibitory, respectively, is generally valid based on their ion selectivity, the direction of a channel-mediated ion current is ultimately determined by a set of additional biophysical conditions e.g. ion gradients and the membrane potential [39,82,83] which were not considered in our work; 4) neurotransmission types other than ligand-gated ionotropic chemical signaling (e.g. G-protein coupled, monamine or neuropeptide) were excluded to avoid mixing different layers of neurotransmission (i.e. extrasynaptic, slow-scale, neuromodulatory transmission) [47,74,75]; 5) even in the case of ionotropic receptors, there are likely a number of additional ligand-gated ion channels that are still uncharacterized [84].

Although the strength of prediction of our work is generally acceptable (>70%), as new data of connectivity and gene expression emerge, our method can be used to provide more accurate predictions of synaptic polarity.

Within the scope of our aims and subject to the limitations discussed above we predicted synaptic polarities in the ionotropic chemical synapse network of *C. elegans* using expression data, for the first time. We developed and applied a novel method that combines connectivity with presynaptic and postsynaptic gene expression data and made its tool available for users at the website http://EleganSign.linkgroup.hu/. Amongst ionotropic chemical synaptic connections, balance of excitatory and inhibitory connections similar to other real-world networks can be well approximated only if one considers both pre- and postsynaptic gene expression, a concept that was lacking from previous work. Our method opens a way to include spatial and temporal dynamics of synaptic polarity that would add a new dimension of plasticity in the excitatory:inhibitory balance. When sufficient data is available, our polarity prediction method can be applied to any neuronal (and as a concept non-neuronal) network.

## Methods

### Description of *C. elegans* connectome data

Connectome reconstruction of the adult hermaphrodite worm published by WormWiring (http://wormwiring.org) consists of 3,638 chemical connections and 2,167 gap junctions, connecting 300 neurons (the two canal-associated neurons, CANL and CANR remained isolated in this reconstruction, and therefore were omitted from the connectome-related analyses). Each of the connections has 4 attributes: the presynaptic neuron, the postsynaptic neuron, the type of the connection (chemical or electrical), and the number of synapses. The chemical connections subset consisting of 20,589 synapses connecting 297 neurons was used in our work (the sensory neuron pair PLML and PLMR, and the pharyngeal neuron M5 is isolated in the chemical synapse network). Additionally, two other connectome reconstructions–both covering a smaller number of neurons and synapses–were used for validation (S5 Table).

### Description of gene expression data and processing

Initially, neuronal binary gene expression data was obtained from a previous publication based on Wormbase [21]. This was extended with data of neuronal neurotransmitter [10,19,20,61] and receptor [24] expression, and with expression data from the recently published CenGen database [23] after transformation to binary information (S1 Text, S6 Table, and S7 Data). For receptor expression scoring, only the genes coding postsynaptic ionotropic receptor subunits were evaluated according to the six functional classes based on their suggested neurotransmitter ligand (glutamate, acetylcholine or GABA) and putative ion channel type. Cation and anion channel genes were categorized as excitatory and inhibitory, respectively. Expression of one or more genes in a functional class was considered positive.

### Prediction of synaptic polarities

Polarities were predicted for connections based on presynaptic neurotransmitter and postsynaptic receptor expression data, using nested logical and conditional formulas. In case of the method referenced as the *NT+R method* throughout the paper, synapses were predicted as *excitatory* or *inhibitory* if only cation channel or only anion channel receptor genes matched the presynaptic neurotransmitter, respectively; *complex* if both types of receptor genes matched; and *unpredicted* if no receptor gene matched. Alternative prediction methods were used according to the following rules. *NT-only method*: synapses were predicted *excitatory* or *inhibitory* if the presynaptic neurotransmitter was acetylcholine and/or glutamate or GABA, respectively; *complex* if acetylcholine and/or glutamate and GABA; and *unpredicted* if the neurotransmitter was none of these. *R-only method*: synapses were predicted *excitatory* or *inhibitory* if the postsynaptic receptor genes expressed were only cation channel or anion channel coding, respectively; *complex* if both types of ion channel receptor genes were expressed; and *unpredicted* if no receptor gene was expressed. Exact formulas are available in Supplementary Data.

### Software and data

Data were processed and predictions were made using Microsoft Excel (ver. 16.32) and R (RStudio 1.1.456) using standard packages.

### Supporting information

**S1 Fig. Proportions of predicted synaptic polarities in alternative *C. elegans* neuronal networks.** Predictions were made based on the neurotransmitter and receptor gene expression patterns of the presynaptic and postsynaptic neurons, respectively (*NT+R method*, see Methods). Red, blue, and grey colors mark inhibitory, excitatory, and complex polarities, respectively. **(A)** Excitatory-inhibitory balances in alternative networks of the WormWiring connectome reconstruction. Bars from top to bottom: 1. synapse weighted network for comparative purpose (same as in Fig 3A); 2. weak links (defined by synapse number of 1) deleted [3]; 3. links connecting any of the pharyngeal nervous system neurons deleted. The rationale is that many previous work analyzed the connectome without the pharyngeal nervous system [2,85]; 4. unweighted network. **(B)** Predicted synaptic polarities for two connectome reconstructions other than Wormwiring, covering a variable number of neurons and synapses [2,3] (S5 Table). In summary, excitatory:inhibitory sign-balance ratios were similar in all cases, ranging between 3.1–4.1. Source data are provided in S1, S4 and S5 Data.
(TIF)

**S2 Fig. Separate representation of the locomotion subnetwork of *C. elegans*.** Figure is a split network representation of Fig 4B. Edges represent excitatory (blue), inhibitory (red), or complex (black) chemical connections. Edges are weighted according to synapse number. The shape of vertices (Δ,○,◇) represent the modality (sensory, inter, motor, respectively) of neurons. **(A)** Head circuit neurons. **(B)** Ventral nerve cord motor neurons. Colors as in Fig 4B.
(TIF)

**S1 Data. Prediction of synaptic signs based on neurotransmitter and receptor expression data.**
(XLSX)

**S2 Data. Prediction of synaptic signs based on neurotransmitter expression data.**
(XLSX)

**S3 Data. Prediction of synaptic signs based on neurotransmitter receptor expression data.**
(XLSX)

**S4 Data. Prediction of synaptic and edge signs based on neurotransmitter and receptor expression data in different subnetworks.**
(XLSX)

**S5 Data. Prediction of synaptic signs based on neurotransmitter and receptor expression data (alternative connectome reconstructions).**
(XLSX)

**S6 Data. Summary of predictions as in S1, S4, and S5 Data after exclusion of the RNA-seq dataset.**
(XLSX)

**S7 Data. Utilization and curation of data sources.**
(XLSX)

**S1 File. High-resolution vector graphic version of Fig 5.**
(PDF)

**S1 Table. Channel type (cation or anion) of ionotropic neurotransmitter receptor genes of *C. elegans*.** The *Caenorhabditis elegans* genome contains 62 ionotropic postsynaptic receptor genes for glutamate, acetylcholine, and GABA. In this table genes are listed in alphabetic order, and the type of channel (+ for cation channel,–for anion channel) is presented with relevant reference. The 42 genes that were expressed postsynaptically in at least one neuron in our database (marked **bold**) have been validated for being cationic or anionic.
(DOCX)

**S2 Table. Distribution of neurons according to the number of ionotropic neurotransmitter receptor genes expressed.** Neuronal gene expression database was compiled from available datasets and manually curated (Methods). Genes encoding ionotropic receptors for glutamate,

acetylcholine or GABA were grouped according to the type of ion channel (cation or anion), i.e. whether being excitatory or inhibitory. Bold numbers represent number of neurons expressing certain numbers of excitatory and/or inhibitory receptor genes. 63 neurons express only cation-channel receptor genes (green), while 48 neurons express only anion-channel receptor genes (red). 151 neurons express a mixture of cation- and anion-channel receptor genes (grey). Source data is in S1 Data.
(DOCX)

**S3 Table. Validation of results with a previous synaptic polarity prediction paper.** Predicted polarities from our results (S1 Data) were compared to the polarities predicted by Rakowski and Karbowski, 2017, for the locomotion circuit of the *C. elegans* connectome. Each cell represents a connection between the named source and target neuron. Green colored cell means that the predicted polarity was the same in both cases. Orange colored cell means that the predicted polarity was different with the two methods. 456 of the 652 synapses (70%) were predicted the same.
(DOCX)

**S4 Table. Validation of predictions with previously published experimental results.** Predicted polarities from our results (S1 Data) were individually compared to previously published experimental data. Each row represents a single connection. If validated ("Yes" in column "Validated?"), the *NT+R* predicted polarity equals the reference polarity. Partial validation means that the predicted and/or reference polarity was complex or uncertain.
(DOCX)

**S5 Table. Comparison of three chemical synapse connectome reconstructions.** The three most complete connectome reconstructions of C. elegans, namely the WormWiring (http://wormwiring.org), as well as published by Varshney et al., 2011, and Cook et al., 2019, have fundamental differences in their coverage of chemical connections and synapse numbers.
(DOCX)

**S6 Table. Manual curation and edits.** Bulk gene expression data was updated manually according to literature data. Green and red text shows receptor gene additions and deletions, respectively, to (from) specified neuron groups. Blue text shows neurotransmitter expression additions. *acc-4* and *lgc-46* genes were excluded to avoid false predictions because of literature evidence supporting a presynaptic localization rather than postsynaptic. All neurons of a neuron group were updated unless specified otherwise.
(DOCX)

**S1 Text. Gene expression.**
(DOCX)

## Acknowledgments

We thank members of the LINK network science group (http://linkgroup.hu) for their helpful comments.

## Author Contributions

**Conceptualization:** Bánk G. Fenyves, Peter Csermely.

**Data curation:** Bánk G. Fenyves, Gábor S. Szilágyi.

**Formal analysis:** Bánk G. Fenyves, Gábor S. Szilágyi.

**Funding acquisition:** Bánk G. Fenyves, Csaba Sőti, Peter Csermely.

**Investigation:** Bánk G. Fenyves, Gábor S. Szilágyi.

**Methodology:** Bánk G. Fenyves, Gábor S. Szilágyi.

**Project administration:** Peter Csermely.

**Software:** Zsolt Vassy.

**Supervision:** Csaba Sőti, Peter Csermely.

**Visualization:** Bánk G. Fenyves.

**Writing – original draft:** Bánk G. Fenyves.

**Writing – review & editing:** Bánk G. Fenyves, Gábor S. Szilágyi, Csaba Sőti, Peter Csermely.